# Aspect category learning and sentimental analysis using weakly supervised learning

Kalpa Subbaih[a, b,*], Bharath Kumar Bolla[c]

*[a] UpGrad Education Private Limited, Mumbai, India*
*[b] Liverpool John Moores University, UK*
*[c] Salesforce, Hyderabad, India*
*\* Corresponding author*

**Abstract**

The surge of e-commerce reviews has presented a challenge in manually annotating the vast volume of reviews to comprehend their underlying aspects and sentiments. This research focused on leveraging weakly supervised learning to tackle aspect category learning and the sentiment classification of reviews. Our approach involves the generation of labels for both aspects and sentiments, employing the Snorkel framework of WSL, which incorporates aspect terms, review sentiment scores, and review ratings as sources of weak signals. This innovative strategy significantly reduces the laborious labeling efforts required for processing such extensive datasets. In this study, we deployed hybrid models, namely BiLSTM, CNN-BiLSTM, and CNN-LSTM, which harness multiple inputs, including review text, aspect terms, and ratings. Our proposed model employs two distinct loss functions: Binary Cross Entropy with Sigmoid Activation for Multi-Label Classification, enabling us to learn aspect Labels such as Quality, Usability, Service, Size, and Price, and Categorical Cross Entropy with Softmax Activations for Multi-Class Classification. Subsequently, we meticulously evaluate the performance metrics of these three implemented models, including Macro F1 score and Macro Precision. CNN & Bi-LSTM model attained 0.78 and 0.79 F1 scores on aspect and sentiment identification, respectively. The outcomes of this research are poised to make a substantial contribution to e-commerce platforms, offering an efficient and automated means to label and analyze vast troves of user reviews.

\* Corresponding author, *E-mail address:* bolla111@gmail.com

## 1. Introduction

E-commerce has redefined the retail landscape, leveraging the internet to facilitate buying and selling. Central to this model is understanding and utilizing customer feedback, with 88% of customers making purchase decisions based on online reviews [1]. The surge in customer reviews has made it challenging to manually analyze feedback using traditional databases, giving rise to Big Data solutions like Hadoop for storage and Natural Language Processing (NLP) for understanding user sentiments.

Sentiment Analysis, a crucial application of NLP, classifies sentiments at document, sentence, and aspect levels. For instance, in the user review, "The watch is too good but is very costly," the 'product quality' aspect has a positive sentiment, while the 'price' aspect carries a negative sentiment. This form of sentiment analysis often relies on Lexicon-based approaches, Machine Learning, and, more recently, Deep Learning. Deep neural networks such as Bert and BiLSTM with Attention have shown impressive results for Sentiment Classification [1].

However, manual labeling of many e-commerce reviews for sentiment classification is a significant challenge. This issue has drawn research interest in weakly supervised learning (WSL), which uses noisy labeled data as weak signals to label many unlabeled data, thus reducing the laborious effort required for supervised learning [2]

Our study aims to harness WSL for Aspect Category learning and overall review-level sentiment classification using customer ratings as a weak signal. This approach presents a novel avenue to understand customer sentiment, with the advantage of manually reducing the time and effort required to label aspects in the reviews [3].

The current research focuses on understanding key aspects like product quality, usability, size, service, and price, classifying review sentiments as positive, negative, or mixed, with a scope limited to e-commerce review data. The framework aims to streamline Aspect Category learning and sentiment classification using WSL by deploying Deep Learning models. This model's successful implementation will substantially contribute to e-commerce platforms, providing an efficient, automated method to label and analyze vast amounts of user reviews and ratings, paving the way for more informed business growth strategies. The objectives of the study are as follows:

- To manually label the Aspect Category (herein referred to as AC) and Aspect Terms for a subset of reviews.
- To apply enhanced pre-processing and utilize WSL to acquire AC and Sentiment labels for unlabelled data.
- To devise an improved architecture for AC learning (herein referred to as ACL), incorporating Aspect Terms as a weak signal and enhancing Review Sentiment Classification by using Review Rating as a weak signal.
- To conduct a comparative analysis of predictive models and evaluate their performance using the Amazon reviews sentiment analysis dataset.

## 2. Literature Review

User reviews in e-commerce hold critical importance for business growth. By analyzing customer feedback, product owners can glean insights into product aspects such as quality, service, price, size, and usability, enabling them to align their offerings more accurately with customer preferences and enhance profitability [4]. Sentiment analysis is a powerful tool used to decipher these customer sentiments, employed in various applications such as social media monitoring, market and competitor research, employee review analysis, and product analytics.

Sentiment analysis techniques employ various tools and methodologies to ascertain subjective information like opinions and emotions from user feedback. These techniques are aimed at classifying user sentiments into positive, negative, or mixed categories, a process crucial to understanding user experiences and preferences in e-commerce.

Before any sentiment analysis can occur, user reviews must undergo pre-processing, which involves sentence segmentation, lower-case conversion, tokenization, part-of-speech tagging, stop word removal, regular expressions for removing unwanted elements, stemming, and lemmatization. These techniques facilitate more accurate modeling of user reviews [4]. Furthermore, to represent words as vectors for further processing, techniques such as Bag of Words, TF-IDF, and word embeddings like Word2Vec, Glove, fast text, ELMO, and BERT are employed [5].

*2.1. Sentiment analysis*

Sentiment analysis occurs at different levels, including document, sentence, and entity or aspect levels. Each level carries unique insights and understanding of user sentiment toward various product or service aspects [6]. The two broad approaches to sentiment analysis are Lexicon-based and Machine Learning-based. The former relies on dictionary and corpus-based methods, while the latter leverages supervised machine learning algorithms such as Naïve Bayes, Support Vector Machines, and Maximum Entropy [7]. Recently, deep learning models, including Convolutional Neural Networks, Recurrent Neural Networks, and Deep Neural Networks, have been increasingly used for sentiment analysis due to their superior performance in processing high-dimensional data and handling overlapping sentiments [8].

Recent research into sentiment analysis has spanned both review-level and aspect-level analyses. At the review level, studies have leveraged machine learning, deep learning, and newer models like BERT for sentiment classification across diverse domains, including restaurants, online courses, and product reviews on platforms like

Amazon [9] [10] [11] [12]. For aspect-level sentiment analysis, researchers have utilized techniques like SVM [13] [14], Latent Dirichlet Allocation (LDA) with Word2Vec [15], topic modeling [16], BiLSTM with attention mechanisms [17] [18], autoencoder [19] and Convolutional Neural Network with self-attention [20], as well as the BERT model [21]. This has enabled a detailed understanding of user sentiments concerning different aspects of products or services, leading to potential improvements.

*2.2. Weak supervision*

The challenge of manually labeling large amounts of data for sentiment classification has led to the increasing use of WSL methods. These methods have been deployed to extract sentiment from review-level and aspect-level data, often in combination with traditional models such as Naïve-Bayes, Convolutional Neural Networks (CNN), and Bi-directional Long Short-term Memory (BiLSTM) [22]. Additionally, ratings provided by users, although possibly noisy, have been used as weak supervision signals to learn suitable embedding spaces for review sentiment classification [3]. However, no approach thus far has leveraged WSL to extract both aspect categories and overall review sentiments while considering user ratings. This would reduce the need for extensive data labeling and better capture the user's sentiment, who may provide different aspect-level comments and overall ratings. Such understanding would allow eCommerce businesses to improve user experience on specific aspects. Therefore, it's critical to consider review text and review ratings and employ WSL for comprehensive and efficient sentiment analysis.

*2.3. Snorkel*

Snorkel, initiated at Stanford in 2016, provides a solution for weak supervision by creating weak labels through Python scripts or inexpensive annotators to train models. It employs a generative model to denoise data without requiring ground truth labels, learning from statistical distributions, and data correlations [23] . Snorkel's architecture integrates data programming, enabling the creation of label functions using rules, heuristics, and external knowledge bases. This results in a set of probabilistic labels that a discriminative model uses to generalize beyond the label functions, improving unseen data coverage. The system has been successfully applied in real-world scenarios such as product classification and real-time event classification [24] [23].

## 3. Research Methodology

*3.1. Dataset Description*

The dataset utilized for this research is the "Amazon Reviews for Sentiment Analysis" from Kaggle [25], which includes millions of customer reviews and corresponding star ratings. Ratings are label_1 (1- and 2-star reviews) and label_2 (3- and 4-star reviews). Each entry in the dataset contains a label and a corresponding review text, with 3,600,000 reviews collected for products sold by Amazon. This dataset is not a standard dataset to benchmark the performance of various models. Train and test data are not separated in this data. This is a real-time uncleaned review data.

*3.2. Solution*

Our research aims to establish a practical approach to ascertain a review's AC and analyze its sentiments. Our proposed solution unfolds in two stages. In the first stage (Fig 1), the annotated reviews and aspect-related keywords are used to generate the weak labels in a probabilistic framework for learning the AC of unlabeled reviews. The second stage (Fig 1) harnesses sentiment scores derived from Sentiment Vader and user ratings as weak labels for sentiment classification. Sentiment Vader is a rule-based tool that scores positive, negative, and neutral sentiments. Five hundred observations were manually labelled with aspects and review sentiment to create test data.

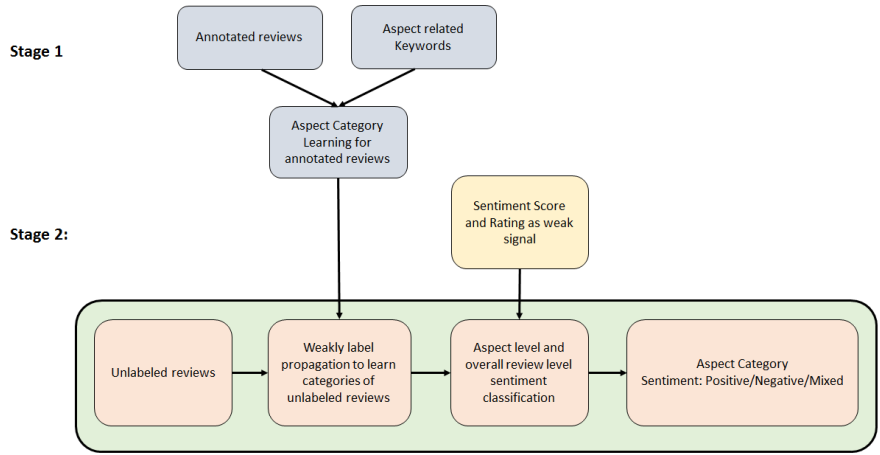

Fig. 1. Process of utilizing WSL for aspect and sentiment rating reviews.

## 3.3. Data preprocessing

The data processing involves splitting the review text and review ratings into two columns, converting the text to lowercase, removing digits, URLs, and punctuations, and applying stop word removal and lemmatization to refine the content. Word Cloud is used to visualize keywords in the review text. The dataset is filtered to include only product-related reviews, excluding those for movies, songs, or books, focusing on aspect categories like Usability, Price, Size, Service, and Quality. Some reviews are manually annotated with aspect categories, aspect terms, and sentiment labels ("Positive," "Negative," "Mixed"). These labeled reviews serve as a learning basis for unlabeled reviews. The review text is tokenized to convert into numerical form using text-to-sequence, and text-to-matrix is applied to the aspect terms for ACL. Label Encoder converts the AC column numerically, and one-hot encoding is applied to create separate columns for each category, ensuring equal weightage for all values in the Aspect Category column.

Table 1. Aspect categories and corresponding aspect terms

| Aspect Category | Aspect Terms |
|---|---|
| **Price Terms** | cost, price, inexpensive, investment, money, penny, pay, cheap, spent, pricy, priced, expensive, cheaper, costs, cheapest, free, paid, dollar, overpriced, bucks, pricing, budget, tax, Money, $, expense, costly, fee |
| **Size Terms** | size, fits, heavy, sizes, chart, smaller, large, feet, big, fit, longer, small, tiny, width, thin, taller, tight, small inch, skinny, hefty, long, xl, length, ft, inches, measurement, stretched, medium, large, sized, smaller size |
| **Service Terms** | manual, instructions, contact, seller, shipping, return, arrived, cardboard box, box, packaging, packaged, date, contacted, response, refund, apologized, trust, duplicate, delivered, advertisement |
| **Quality Terms** | quality, broken, tore, lasts, inferior, solid, brass, scraped, smells, delicate, plastic, stiff, tolerate, textured, chinsy, blunt, sharp edges, sharp, waterproof, soft, smell, smooths, broke, poor, textureline, fabric |
| **Usability Terms** | useful, functions, flexibility, using, useless, work, uncomfortable, performance, use, works, job, functionality, designed, usable |

*3.4. Weak Label Generation*

Label functions are leveraged to categorize sentiments and aspect categories of reviews, using ratings and sentiment terms as weak signals for sentiment classification. Snorkel's generative and discriminative models are used for labeling the reviews. As for AC labeling, aspect terms, identified from manually annotated reviews, are used as a weak source for learning. If words in the review match any list of aspect terms related to Usability, Price, Size, Service, or Quality, the corresponding AC is assigned (Table 1); otherwise, the function excludes assigning a label as described in Fig 2. Given that ACL constitutes a multi-label classification challenge, the 'MajorityLabelVoter' model within Snorkel is applied to acquire the AC labels. The predict_proba function within 'MajorityLabelVoter' furnishes all classes with a probability score exceeding 0. The parameter "Cardinality" is set at five to accommodate the five aspect classes.

For sentiment labeling, positive, negative, or mixed labels are assigned based on the polarity score of sentiment terms and ratings. The acquisition of sentiment labels involves utilizing the sentiment score extracted from the Sentiment Vader library, coupled with ratings as a less robust signal. The Sentiment Vader library, harnessed from the nltk package, employs rule-based sentiment analysis to furnish sentiment polarity. A compound score exceeding 0.05 indicates positive sentiment, while a compound score below -0.05 is construed as indicative of negative sentiment. For multi-classification labelling, the LabelModel within Snorkel is configured with a cardinality parameter of three to accommodate the labels Positive, Negative, and Mixed. Sample output of labelled data is shown in Table 2.

---

**Algorithm 1:** Aspect label functions to generate labels for aspect categories using Snorkel library

1: **input**   :*R*: the Review Text, *AD*: dictionary of Aspects *A* as keys and list of Aspect Terms **AT** as
            values for each key *A*
2: **output :** the list of aspect labels *L* for each review
3: **for** each review *r* ∈ *R* **do**
4:     **for** each aspect *A* ∈ *AD* **do**
5:         initialize count=0
6:         **for** each aspect term *at* ∈ *AT* in *AD[A]* **do**
7:             **if**  *at* in *R*
8:                 increase the count by 1
9:         **end for**
10:        **if** count > 1 **then**
11:            add aspect *A* to the list of label *L*
12:        **else**
13:            add abstain to the list of label *L*
14:        **end if**
15:    **end for**
16: **return** *L*

Fig. 2. Process of utilizing WSL for aspect and sentiment rating reviews.

Table 2. Unlabelled reviews are labeled using Snorkel labeling functions

| Review Text | Snorkel Aspect Category | Snorkel Sentiment Label |
|---|---|---|
| great product. The peerless universal rolling | Price, Quality, Service, Size, Usability | Positive |
| no no no ; this item will smell for about 2 weeks | Quality | Negative |
| Locking gas cap; I ordered a locking gas cap a .. | Service, Size, Usability | Mixed |
| can't live without it; Frederic shakai shea bu | Quality, Size, Usability | Positive |
| don't waste your money; I'm a fairly intelligent | Price, Size, Usability | Negative |

*3.5. Hybrid Models on Weak Labels*

To derive both the AC and Review Sentiment, a hybrid model is created that utilizes multiple inputs, including Review text, Aspect Terms, and Rating. Glove is used to convert Review text into a vector representation. Various architectures were experimented with, including deep neural networks such as CNN, LSTM, and BiLSTM, which have shown promising results in the literature. Three hybrid models were used in the research. BiLSTM hybrid model takes review text and applies BiLSTM architecture (Fig 3.). In the CNN-LSTM hybrid model (Fig-4), LSTM's output, having learned the historical information from lengthy review texts, serves as input for CNN to identify local features. Another architecture, BiLSTM with CNN, combines BiLSTM for semantic extraction from the review text and CNN for keyword extraction (Fig. 5). Dropout and L2 Regularization techniques were implemented to prevent overfitting in the model. Rectified Linear Unit (ReLU) [26] activation functions were employed within the model's hidden layers. For Aspect Learning, Sigmoid activation was applied [27], while for Sentiment classification, Softmax activation was utilized [28]. These measures collectively contribute to a more robust and generalizable model, reducing the risk of overfitting and enhancing its predictive capabilities.

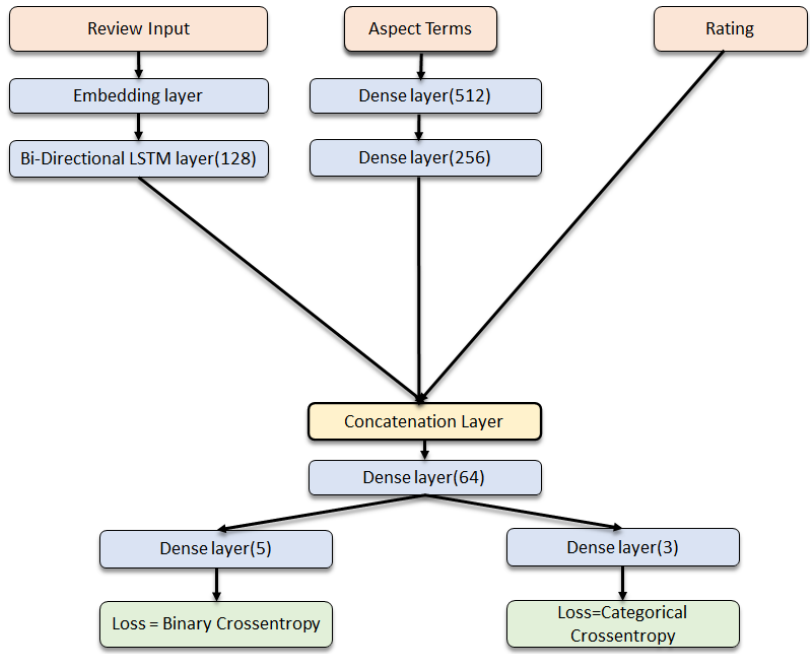

Fig. 3. BiLSTM hybrid model for Aspect Category and sentiment classification

The models developed for ACL and sentiment classification on the Amazon Review for Sentiment Analysis dataset will be evaluated and compared using Macro F1 Score. Other metrics are also analyzed, including Macro Precision, Macro Recall, Micro F1 Score, Micro Precision, Micro Recall, and Hamming Loss. The model with the highest Macro F1 Score, which effectively penalizes imbalanced data, will be chosen as the final model. Some predicted aspect categories and sentiment classifications will be manually checked to ensure the predictions align with real-world business understanding. The hybrid model will utilize two loss functions: Binary Cross Entropy (with Sigmoid Activation) for Multi Label Classification to learn the Aspect Labels Quality, Usability, Service, Size, and Price; and Categorical Cross Entropy (with Softmax Activation) for Multi Class Classification to learn the sentiments Positive, Negative, and Mixed. Relu Activation is used in hidden layers to prevent gradient vanishing and aid in faster neural network convergence.

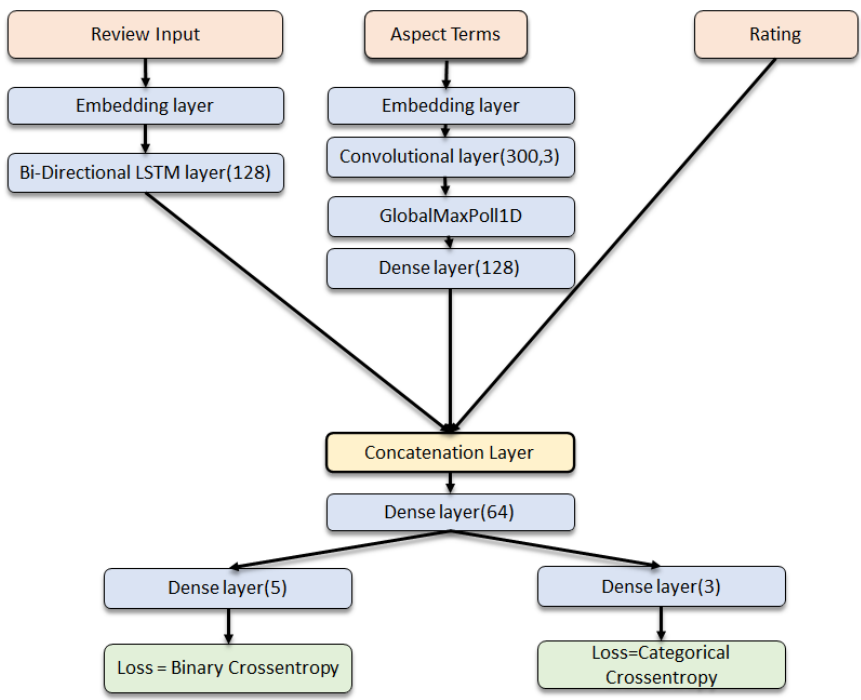

Fig. 4. CNN-BiLSTM hybrid model for Aspect Category and sentiment classification

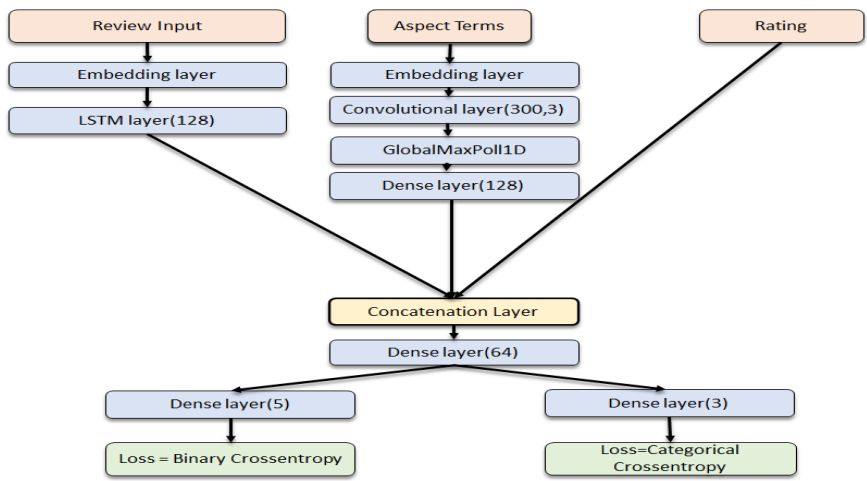

Fig. 5. CNN-LSTM hybrid model for Aspect Category and sentiment classification

# 4. Results

This section discusses the results obtained from the Snorkel labeling model on test data and hybrid models trained on Snorkel labeled data.

## 4.1. Aspect Category snorkel labeling model output

This section discusses the performance of the Snorkel labeling function for AC labels. Table 3 summarizes the function's output, indicating the unique labels generated, the percentage of the dataset it covers, the overlap with other labeling functions, and any disagreements with different functions. Notably, the labeling function for the "Usability" category has high coverage. This observation aligns with our earlier data exploration, which showed a large number of reviews related to usability.

Table 3. Aspect labeling functions output on train data

| Labeling Function | Polarity | Coverage | Overlaps | Conflicts |
|---|---|---|---|---|
| lf_price | [0] | 0.304300 | 0.27 | 0.27 |
| lf_size | [3] | 0.292553 | 0.27 | 0.27 |
| lf_service | [2] | 0.358599 | 0.30 | 0.30 |
| lf_quality | [1] | 0.368720 | 0.33 | 0.33 |
| lf_usability | [4] | 0.638372 | 0.49 | 0.49 |

Next, Table 4 compares the model's performance on the test data. The Macro and Micro metrics are pretty similar. However, the precision value is lower than the recall value, suggesting that the model has a higher rate of false positives.

Table 4. Snorkel Majority label model output on test data

| Macro F1 | Macro Precision | Macro Recall | Micro F1 | Micro Precision | Micro Recall | Hamming Loss |
|---|---|---|---|---|---|---|
| 0.86 | 0.76 | 0.98 | 0.86 | 0.77 | 0.98 | 0.11 |

## 4.2. Sentiment category snorkel labeling model output

This section analyzes the Snorkel labeling function's results for Sentiment labels. Table 5. shows no overlaps or conflicts in the labeling process, a desirable quality for a multi-label classification task. Among the labels, "Positive" functions cover a larger portion of reviews, while "Mixed" functions cater to fewer, reflecting that mixed sentiments are less prevalent in reviews.

Table 5. Sentiment labeling output on train data

|  | Polarity | Coverage | Overlaps | Conflicts |
|---|---|---|---|---|
| lf_positive | [1] | 0.421247 | 0.0 | 0.0 |
| lf_negative | [0] | 0.306664 | 0.0 | 0.0 |
| lf_mixed | [2] | 0.272089 | 0.0 | 0.0 |

The performance metrics for the predicted labels on the test data are exhibited in Table 6. Both Macro and Micro metrics yield similar results. However, precision is lower than recall, suggesting a higher rate of false positives in the model's predictions.

Table 6. Sentiment output on test data

| Macro F1 | Macro Precision | Macro Recall | Micro F1 | Micro Precision | Micro Recall | Hamming Loss |
|---|---|---|---|---|---|---|
| 0.89 | 0.89 | 0.92 | 0.92 | 0.92 | 0.92 | 0.05 |

## 4.3. Hybrid model output

This section reviews the performance metrics for the three hybrid models experimented with, namely BiLSTM, CNN-BiLSTM, and CNN-LSTM, as tested on the test data. These models were designed to process inputs—Tokenized Review Text, Tokenized Aspect Terms, and Rating—and produce multiple outputs, namely AC and Sentiment labels.

Table 7. Aspect Category metrics results for the Hybrid models

| Hybrid Model | Macro F1 | Macro Precision | Macro Recall | Micro F1 | Micro Precision | Micro Recall | Hamming Loss |
|---|---|---|---|---|---|---|---|
| Bi-LSTM | 0.81 | 0.76 | 0.88 | 0.82 | 0.78 | 0.87 | 0.14 |
| CNN & Bi-LSTM | 0.78 | 0.68 | 0.92 | 0.8 | 0.7 | 0.94 | 0.17 |
| CNN & LSTM | 0.59 | 0.6 | 0.58 | 0.55 | 0.61 | 0.55 | 0.2 |

Table 7. displays the metric evaluation for ACL across all three models. It details Macro F1, Macro Precision, Macro Recall, Micro F1, Micro Precision, Micro Recall, and Hamming Loss scores. The Macro F1 score, which penalizes imbalanced classes more than the Micro F1 score, was chosen to select the best performing model. The Macro F1 scores were 81% for BiLSTM, 78% for CNN-BiLSTM, and 59% for CNN-LSTM, respectively.

11The Sentiment Classification results in Table 8. reveal Macro F1 scores of 72% for BiLSTM, 79% for CNN-BiLSTM, and 64% for CNN-LSTM. Despite being trained for fewer epochs, the BiLSTM model performs comparably or better than the other two. This suggests that with more training, the BiLSTM model has the potential to deliver superior results.

Table 8. Sentiment classification results for the Hybrid models

| Hybrid Model | Macro F1 | Macro Precision | Macro Recall | Micro F1 | Micro Precision | Micro Recall | Hamming Loss |
|---|---|---|---|---|---|---|---|
| Bi-LSTM | 0.72 | 0.78 | 0.75 | 0.76 | 0.77 | 0.77 | 0.15 |
| CNN & Bi-LSTM | 0.79 | 0.8 | 0.81 | 0.83 | 0.84 | 0.84 | 0.1 |
| CNN & LSTM | 0.64 | 0.62 | 0.65 | 0.81 | 0.81 | 0.81 | 0.12 |

This section examined various evaluation metrics pertaining to the AC labels generated by the Snorkel Majority Label Voter model, followed by metrics for sentiment labels predicted by the Snorkel Label Model. Also discussed the performance metrics, including Macro F1 score and Macro Precision, for the three implemented models: BiLSTM, CNN-BiLSTM, and CNN-LSTM.

## 5. Conclusion

The primary objective of our research was to create a hybrid model capable of predicting both the AC and sentiment labels for Amazon product reviews. We introduce a hybrid model designed to predict both the Aspect Categories and Review Sentiments within online product reviews by harnessing the power of WSL. This model is instrumental in reducing the substantial manual labeling efforts typically required when categorizing aspects like Usability, Price, Size, Service, and Quality. Instead, we employ aspect terms as weak signals and a limited set of manually annotated reviews to facilitate aspect learning. Within our hybrid model, we've developed labeling functions for sentiment classification (Positive, Negative, or Mixed) that consider both the review's sentiment score and the review rating as sources of weak signals.

Our initial step involved manually labeling five hundred reviews and identifying aspects such as Usability, Price, Size, Service, and Quality. Subsequently, aspect terms were extracted, and sentiment labels (Positive, Negative, or Mixed) were manually assigned. It should be noted that these sentiment labels were applied exclusively to test data, serving as a valuable tool for evaluation.

In this study, we've assessed the efficacy of numerous models in ascertaining the AC and Sentiment labels for the reviews. A detailed analysis of metrics corresponding to the AC and Sentiment labeling outputs from Snorkel has been carried out. Following this, an exhaustive evaluation of implemented hybrid Models provides a holistic overview of the paper's content. The implications of this research extend to future endeavors in e-commerce review sentiment analysis, particularly within the domain of WSL, providing valuable insights for developing advanced models in this field. One main limitation of this approach is that the sentiment is not assigned to relevant aspects. Aspects and sentiments are dealt with separately. In future studies, we plan to associate sentiment with the corresponding aspect. Increasing the test data size by manually labeling the data will amplify the robustness of the results.

## 6. Contribution of our work

This research endeavor has made a notable contribution to the domain of WSL. Historically, there has been a scarcity of investigations devoted to proficiently mastering AC and review sentiment using WSL techniques, including review text and rating. The essential advantage of this approach lies in its ability to develop a hybrid model encompassing both ACL and review sentiment classification, leveraging labeling functions as sources of weak signals. These functions permit the rapid and efficient labeling of a substantial volume of data with minimal time and effort by leveraging a few annotated data points alongside the labeling functions.

This research has facilitated the implementation of hybrid models for multi-label and multi-class classification. Within this hybrid model, Aspect terms function as weak signals for ACL, complemented by Review Ratings and Review sentiment scores, which serve as weak signals for the comprehensive sentiment classification. The novelty of the hybrid model lies not only in its capacity to provide aspect categories and sentiment labels as outputs using WSL but also in its incorporation of ratings alongside review text for learning review sentiment. This innovative approach offers an efficient solution to tackle the multifaceted challenges of sentiment classification, all while minimizing the laborious manual labeling process.

In a broader context, the insights from this study hold applicability across various use cases, such as identifying key terms in legal text [29] [30], product analysis, aspect extraction and sentiment analysis of any online reviews, identifying the adverse drug reactions [31] and clinical entities in biomedical text [32] , all without the resource-intensive task of manual labeling.